\documentclass{svproc}
\usepackage{url}

\usepackage{amsmath}
\usepackage{amssymb}
\usepackage{hyperref}
\usepackage[dvipsnames]{xcolor}
\usepackage{slashed}
\usepackage{mathrsfs}
\usepackage{braket}
\usepackage{graphicx, xcolor}
\usepackage{caption,subcaption}
\usepackage[makeroom]{cancel}
\usepackage{hyperref}
\begin{document}
\mainmatter              
\title{Exploring the Effect of Chiral Torsion on Neutrino Oscillation in Long Baseline Experiments}
\titlerunning{Effect of torsion in DUNE}  
%
\author{Riya Barick \and Amitabha Lahiri}
\authorrunning{Barick and Lahiri} 
%
\tocauthor{Riya Barick, Amitabha Laihiri}
\institute{S. N. Bose National Centre for Basic Sciences, Salt Lake, WB 700106, India.\\
\email{riyabarik7@gmail.com ; amitabha@bose.res.in}\\
{\em\small Presented by Riya Barick at the XXVI DAE-BRNS HEP Symposium,\\ 19-23 Dec 2024, Varanasi}}

\maketitle              

\begin{abstract}
In curved spacetime, neutrinos experience an extra contribution to their effective Hamiltonian
coming from a torsion-induced four-fermion interaction that is diagonal in mass basis and also
causes neutrino mixing while propagating through fermionic matter. This geometrical quartic
interaction term appears as the modification to the neutrino mass term and significantly influences
both neutrino conversion and survival probabilities. Since this term varies linearly with matter
density, long baseline (LBL) experiments would be a good choice to probe this effect. We put
bounds on torsional coupling parameters and also see the impact of torsion on physics sensitivities in the DUNE experiment.
\end{abstract}
\section{Introduction}
The dynamics of fermions in curved spacetime require a spin connection which has two components, one is torsion free Levi-Civita connection and the other is contorsion, expressed as $\, A_{\mu}{}^{ab}=\omega_{\mu}{}^{ab}+\Lambda_{\mu}{}^{ab}\,.$ Most generally, the contorsion $\Lambda_{\mu}{}^{ab}$ couples chirally and it is non-dynamical thus can be integrated out from the theory, leaving an effective quartic interaction term~\cite{Chakrabarty:2019cau}
\begin{equation}
	-\frac{1}{2}(\sum\limits_i (\lambda^V_{i}\bar{\psi}^i \gamma_a  \psi^i + \lambda^A_{i}\bar{\psi}^i \gamma_a \gamma^5 \psi^i))^2\,,
	\label{quartic-int}
\end{equation}
which is diagonal in the mass basis (NSI in mass basis). Here the $\lambda$'s are non-universal geometrical coupling constants to be determined from experiments. Since this is a fundamental interaction, all fermions experience this effect. If we consider the propagation of neutrinos within the Earth matter at constant density, the interaction with the background is given by~\cite{Chakrabarty:2019cau}
\begin{equation}
-(\sum\limits_{i=1,2,3}(\lambda_{i}^{L}\bar{\nu}_i \gamma_a L\nu_i + {\lambda_{i}^R \bar{\nu}_i \gamma_a R \nu_i} )) \times (\sum_{f=e, u, d}(\lambda_{f}^{V}\bar{f} \gamma_a f + \lambda_{f}^A \bar{f} \gamma_a\gamma^5 f ))\,.
\label{interaction-background}
\end{equation}
The geometrical contribution in the second term is reduced to the weighted number density of background fermions $ \tilde{n} = \sum_{f}\lambda_{f} n_f\,.$ This is analogous to the Wolfenstein effect for weak interactions, but with different coupling constants for different fermions. Also, we have assumed that only left handed neutrinos exist in nature, so the effective contribution to the Hamiltonian is
{$\sum\limits_{i=1,2,3}(\lambda_{i}{\nu}_i^\dagger \mathbb{L} \nu_i )\,\tilde{n}$\,.}
\section{Neutrino Oscillation in Matter}
Neutrino flavor eigenstates $ \ket{\nu_{\alpha}}$, can be expressed in terms of mass eigenstates $\ket{\nu_i}$, as $\ket{\nu_{\alpha}} = \sum_{i}U^*_{\alpha i} \ket{ \nu_i} \,,$ where U is the PMNS matrix. The Schr\"odinger equation in flavor basis is given by~\cite{Barick:2023wxx}
\begin{small}
\begin{equation}
i({d/}{dt})  \ket{\nu_e, \nu_\mu, \nu_\tau} = [2\tilde{\Delta}/L][U^*\mathrm{diag}(0,\tilde{\alpha},1)U^T+\mathrm{diag}(\tilde{A}, 0, 0)] \ket{\nu_e, \nu_\mu,  \nu_\tau} \,,
\label{schrodinger-eqn}
\end{equation}
\end{small}
where $\tilde{\alpha}=\frac{\Delta \tilde{m}^2_{21}}{\Delta \tilde{m}^2_{31} }\,,$ $\tilde{A}=\frac{2 \sqrt{2} G_F n_e E}{ \Delta \tilde{m}^2_{31} }\,\mathrm{and}\,\tilde{\Delta} = \frac{\Delta \tilde{m}^2_{31} L}{4E}$. Torsion induced mass-squared difference $ \Delta \tilde{m}^2_{ij}:=\Delta m^2_{ij} +2 \tilde{n} E \lambda_{ij}$ where  $\tilde{n}=(\lambda_e + 3\lambda_u + 3\lambda_d )n_e$ and $ \lambda_{ij}=\lambda_{i}-\lambda_{j}$. 
In Eq~\ref{schrodinger-eqn}, we have not mentioned the term proportional to the identity matrix, since it has no effect on oscillation. We have solved this equation using perturbation technique assuming $\tilde{\alpha} \, \mathrm{and} \, \sin\theta_{13}$ as small parameters \cite{Barick:2023wxx,Barick:2023qjq} and have calculated neutrino conversion and survival probabilities in presence of torsion, given by
\begin{align}
P_{\mu e} =& {4\sin^2\theta_{13} \sin^2 \theta_{23} \frac{\sin^2(\tilde{A}-1)\tilde{\Delta}} {(\tilde{A}-1)^2}} + {\tilde{\alpha}^2 \cos^2 \theta_{23} \sin^2(2\theta_{12}) \frac{\sin^2( \tilde{A} \tilde{\Delta} )}{\tilde{A}^2}} \nonumber \\
&+ 2 \tilde{\alpha} \sin \theta_{13} \sin{2\theta_{12}} \sin{2\theta_{23}} \cos(\tilde{ \Delta} + \delta_{CP} ) \frac{\sin{ \tilde{A} \tilde{\Delta}}}{\tilde{A}} \frac{\sin{( \tilde{A}-1 ) \tilde{\Delta}}}{\tilde{A}-1} \,, \\
P_{\mu \mu } =& 1-{\sin^2( 2 \theta_{23} ) \sin^2 \tilde{\Delta}} + O (\tilde{\alpha},\sin \theta_{13})  \,.
\end{align}
For antineutrinos, the  corresponding probabilities  can be  obtained by making the replacements
$\delta_{CP} \to -\delta_{CP} ,\quad \tilde{A} \to -\tilde{A}, \quad \text{and} \quad \Delta \tilde{m}^2_{ij} := \Delta m^2_{ij} - 2 \tilde{n} E \lambda_{ij}\,$ in the above formulas.
\section{Results and Discussions}
\begin{figure}[hbtp]
\includegraphics[width=0.45\textwidth]{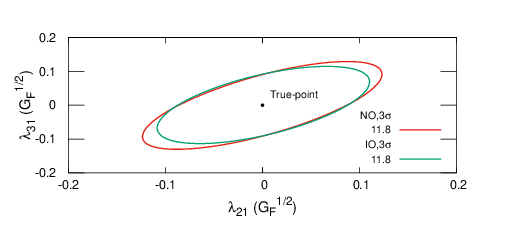}
\hspace{0.5cm}
\includegraphics[width=0.45\textwidth]{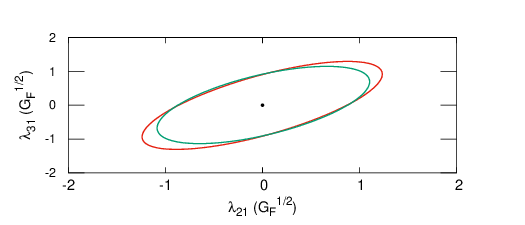}
\caption{\small{Bounds on $\lambda_{21}$ and $\lambda_{31}$ when $\lambda_{e,u,d} = 0.1 $(left) and $0.01$ (right) in DUNE.}}
\label{contour}
\centering
\end{figure}

We have explored the effect of spacetime in the DUNE experiment which is an upcoming long baseline (LBL) experiment, focusing on its far detector with a baseline of 1300 km from Fermilab, USA to SURF, South Dakota. We have simulated DUNE using the General Long Baseline Experiment Simulator (GLoBES)\cite{Huber:2004ka,Huber:2007ji}, a C-based framework, based on the technical design report provided in \cite{DUNE:2021cuw}, taking a total runtime of 13 years (6.5 years $\nu $ + 6.5 years $ \bar{\nu}$). We have modified the GLoBES framework accordingly to incorporate the effects of geometrical four-fermion interaction. To do the statistical analysis we have used the formula $$\chi^2 = 2\sum_i [N^{\rm test}_i-N^{\rm true}_i-N^{\rm true}_i\log(N^{\rm test}_i/N^{\rm true}_i)]\,,$$
where $ N_i$ is the event number in the $i$-th energy bin. In Fig.~\ref{contour}, we have shown the bounds of $ \lambda_{21}$ and $\lambda_{31}$ in DUNE. We have taken the standard interaction (SI) in the true scenario and torsion in the test scenario, with the other neutrino oscillation parameters kept fixed for both.  The neutrino mixing parameters used in this proceeding are based on \cite{Esteban:2020cvm}. For two degrees of freedom $1\sigma\,,2\sigma\,\mathrm{and}\,3\sigma$ correspond to  $\chi^2=2.3\,,6.18\,\mathrm{and}\,11.83$ respectively. It is clear from Fig.~\ref{contour} that the contour for normal mass ordering (NO) is bigger than that of inverted ordering (IO). Consider $\lambda$ in the unit of $\sqrt{G_F}$.
When background torsion $\lambda_{e,u,d} = 0.1 $, we get the following bound $-0.12\leq\lambda_{21}\leq 0.12$ and $-0.13\leq\lambda_{31}\leq 0.13$ while for $\lambda_{e,u,d} = 0.01$, we get $-1.2\leq\lambda_{21}\leq 1.2$ and $-1.3\leq\lambda_{31}\leq 1.3\,,$ at 3$\sigma$ C.L. for NO.
\begin{figure}[hbtp]
\includegraphics[width=0.45\textwidth]{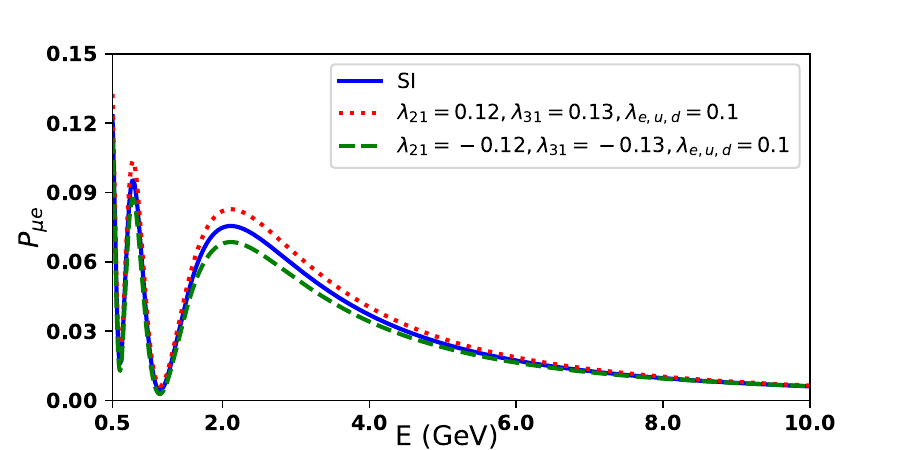}
\includegraphics[width=0.45\textwidth]{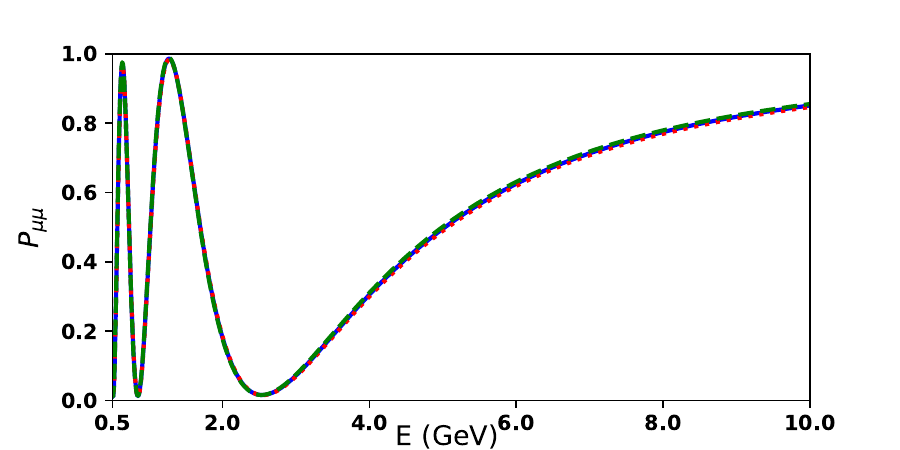}\\
\includegraphics[width=0.45\textwidth]{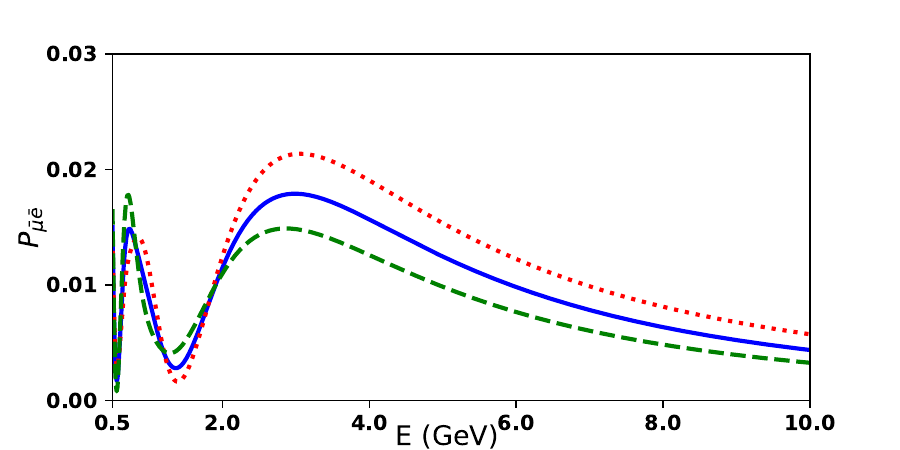}
\includegraphics[width=0.45\textwidth]{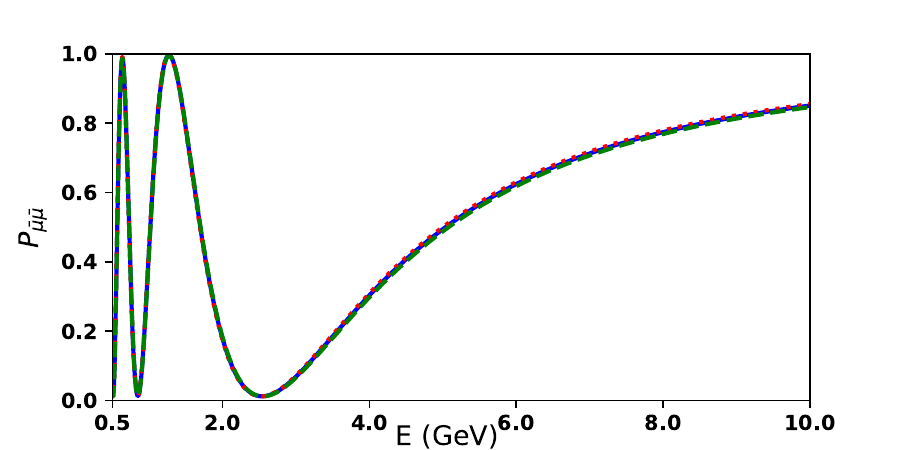}
\caption{\small{Appearance and disappearance probability vs. Energy, for $\nu$ and $\bar{\nu},$ at DUNE.}}
\label{probability-plots}
\end{figure}
Fig.~\ref{probability-plots} shows the effect of geometry on  $\nu_\mu \to \nu_e$ and $\bar{\nu}_\mu\to\bar{\nu}_e$ conversion and $\nu_\mu$ and $\bar{\nu}_{\mu}$ survival probability. Positive $\lambda_{ij}$ enhances the conversion probability while negative $\lambda_{ij}$ suppresses it at the first oscillation peak and their effects are symmetric about the SI plot. The effect of geometry on survival probability is very small in this baseline.
\begin{figure}[hbtp]
\includegraphics[width=0.45\textwidth]{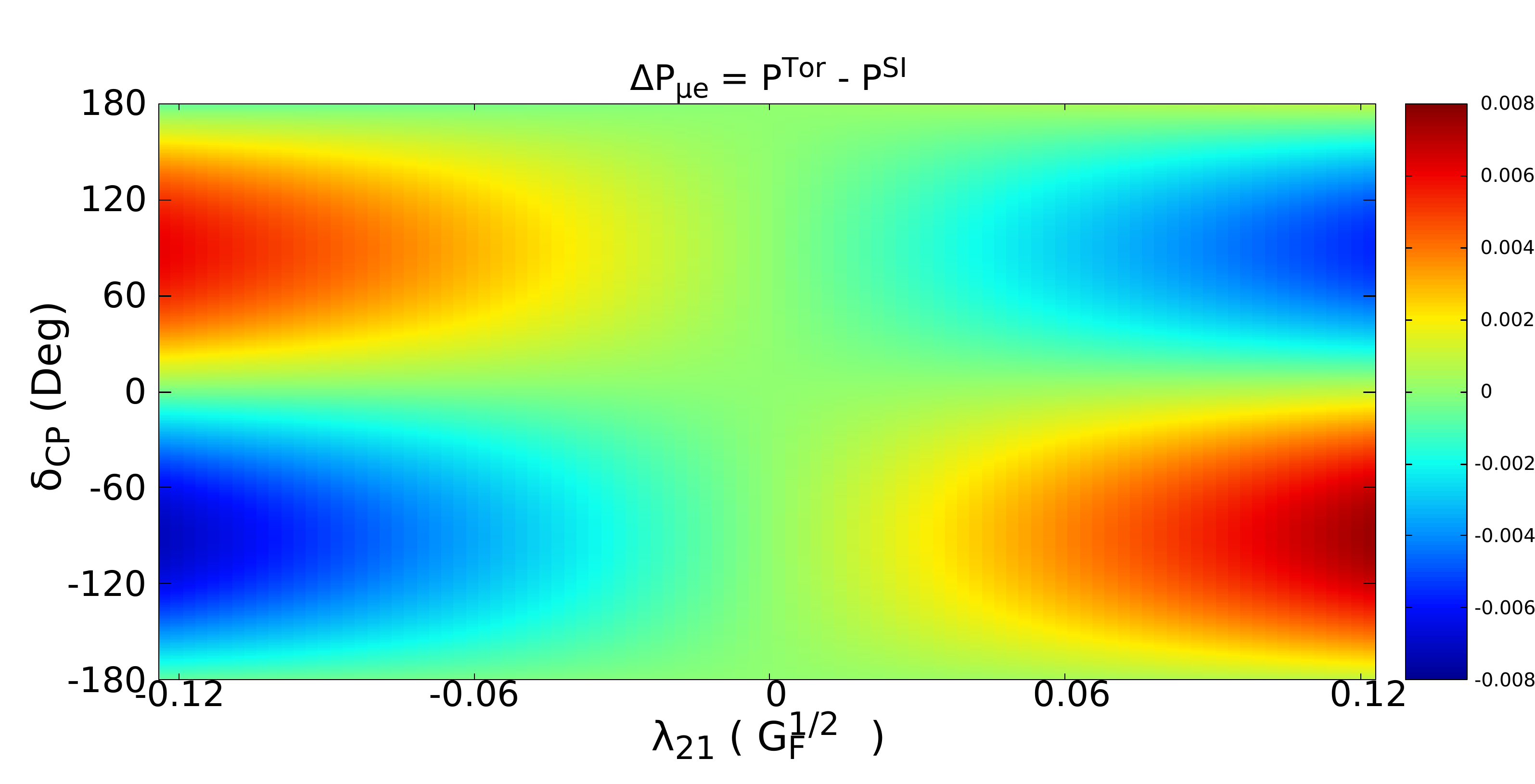} 
\hspace{0.5cm}
\includegraphics[width=0.45\textwidth]{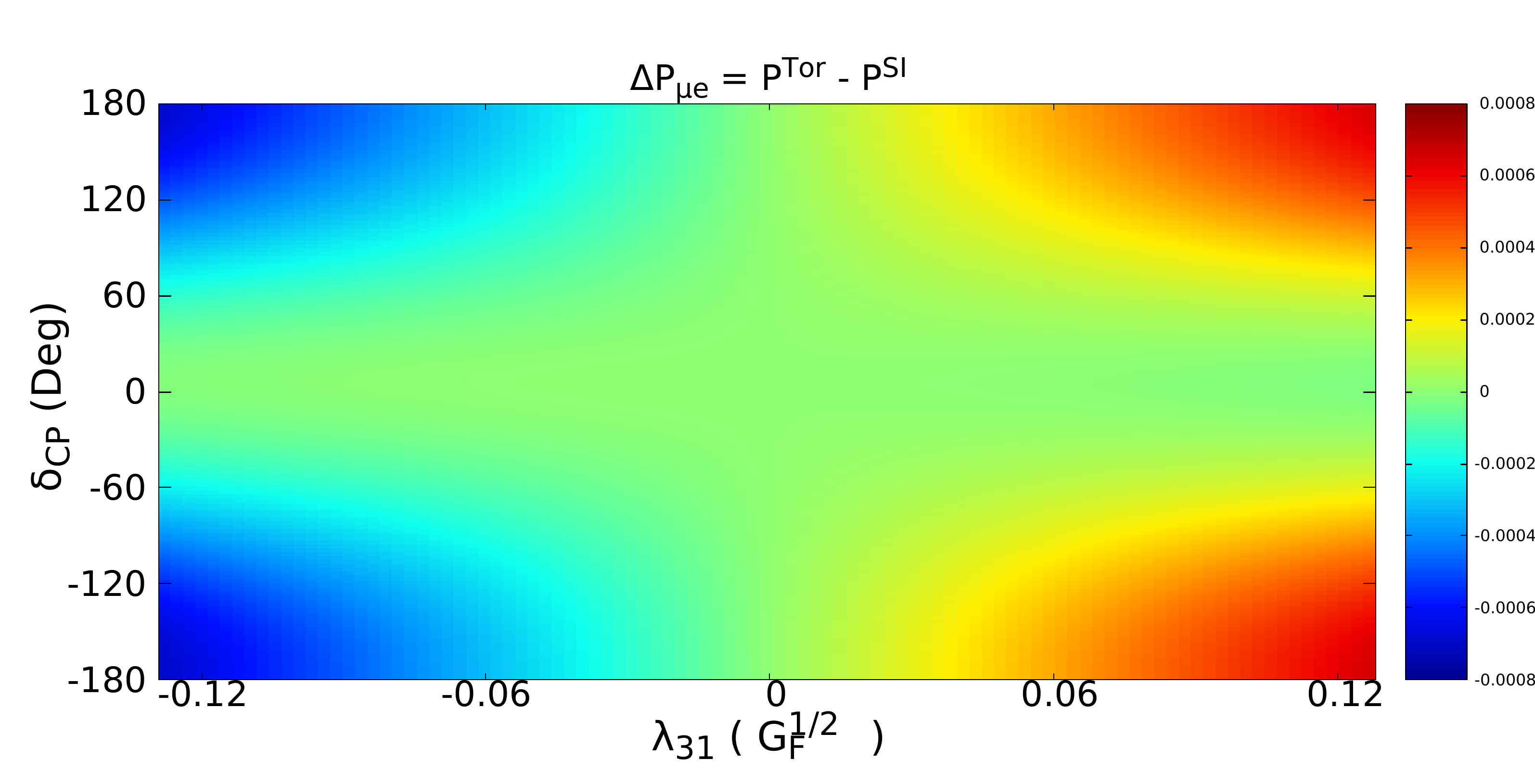}
\caption{\small{$\Delta P_{\mu e} (= P^{Tor}_{\mu e} - P^{SI}_{\mu e})$ in $(\lambda_{ij} - \delta_{CP})$ plane at E = 2.5 GeV for DUNE.}}
\label{color-map}
\centering
\end{figure}
In Fig.~\ref{color-map} we have plotted $\Delta P_{\mu e} = P^{Tor}_{\mu e} - P^{SI}_{\mu e} \,,$ in the $(\lambda_{ij} - \delta_{CP})$ plane at E = 2.5 GeV for DUNE to see the impact of spacetime on $P_{\mu e}$ in the $\delta_{CP}$ parameter space \cite{Sarker:2024ytu}. Here $P^{Tor}_{\mu e}$ is the probability in presence of torsion and $P^{SI}_{\mu e}$ is the standard oscillation probability. We observe that $\lambda_{21}$ and $\lambda_{31}$ have different impacts --- $\Delta P_{\mu e}$ is an order of magnitude larger for $\lambda_{21}$ than for $\lambda_{31}$.
\begin{figure}[hbtp]
\includegraphics[width=0.45\textwidth]{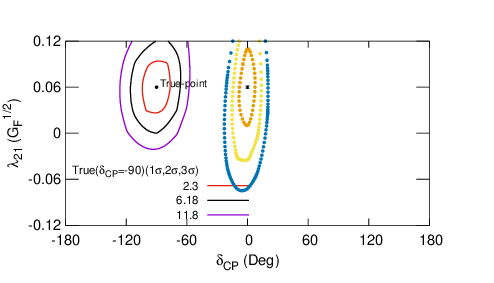} 
\hspace{0.5cm}
\includegraphics[width=0.45\textwidth]{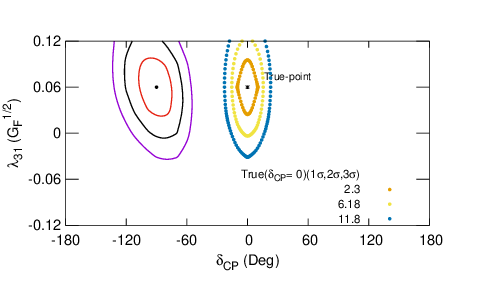}
\caption{$\delta_{CP}$ constraining capability of DUNE in presence of $\lambda_{ij}( \sqrt{G_F})$\,.}
\label{cp-capability}
\centering
\end{figure}
Fig.~\ref{cp-capability} represents the sensitivity of DUNE to constrain $\delta_{CP}$ in presence of torsion. 
We have chosen $\delta^{true}_{CP}=-90^\circ$(CP-violating value) and $0^\circ$ (CP-conserving value) and $\lambda^{true}_{21}=\lambda^{true}_{31}=0.06\sqrt{G_F}\,.$ 
We observe that in presence of $\lambda_{21}\,,$ $\delta_{CP}$ is constrained $\sim [-60^\circ : -120^\circ]$ for $\delta^{true}_{CP}=-90^\circ,$ whereas for $\delta^{true}_{CP}=0^\circ$ we get a relatively better constraint on  $\delta_{CP}\,,$ although it spreads widely along $\lambda_{21}$ range. The right panel shows that, $\delta_{CP}$ constraining is better for $\delta^{true}_{CP}=0^\circ$ than $-90^\circ\,,$but the constraint for $\lambda_{31}$ is comparable.

\textbf{Acknowledgements :} The authors thank I. Ghose, S. Goswami, and S. K. Raut  for fruitful discussions.
%

\end{document}